\documentclass[twoside,twocolumn,9pt]{article}
\usepackage{extsizes}
\usepackage[super,sort&compress,comma]{natbib} 
\usepackage[version=3]{mhchem}
\usepackage[left=1.5cm, right=1.5cm, top=1.785cm, bottom=2.0cm]{geometry}
\usepackage{balance}
\usepackage{mathptmx}
\usepackage{sectsty}
\usepackage{graphicx} 
\usepackage{lastpage}
\usepackage[format=plain,justification=justified,singlelinecheck=false,font={stretch=1.125,small},labelfont=bf,labelsep=space]{caption}
\usepackage{float}
\usepackage{fancyhdr}
\usepackage{fnpos}
\usepackage[english]{babel}
\addto{\captionsenglish}{%
  
}
\usepackage{array}
\usepackage{droidsans}
\usepackage{charter}
\usepackage[T1]{fontenc}
\usepackage[usenames,dvipsnames]{xcolor}
\usepackage{setspace}
\usepackage[compact]{titlesec}
\usepackage{hyperref}

\begin{document}

\pagestyle{fancy}
\thispagestyle{plain}
\fancypagestyle{plain}{
\renewcommand{\headrulewidth}{0pt}
}

\makeFNbottom
\makeatletter
\renewcommand\LARGE{\@setfontsize\LARGE{15pt}{17}}
\renewcommand\Large{\@setfontsize\Large{12pt}{14}}
\renewcommand\large{\@setfontsize\large{10pt}{12}}
\renewcommand\footnotesize{\@setfontsize\footnotesize{7pt}{10}}
\makeatother

\renewcommand{\thefootnote}{\fnsymbol{footnote}}
\renewcommand\footnoterule{\vspace*{1pt}%
\color{black}\vspace*{5pt}} 
\setcounter{secnumdepth}{5}

\makeatletter 
\renewcommand\@biblabel[1]{#1}            
\renewcommand\@makefntext[1]%
{\noindent\makebox[0pt][r]{\@thefnmark\,}#1}
\makeatother 
\renewcommand{\figurename}{\small{Fig.}~}
\sectionfont{\Large}
\subsectionfont{\normalsize}
\subsubsectionfont{\bf}
\setstretch{1.125} 
\setlength{\skip\footins}{0.8cm}
\setlength{\footnotesep}{0.25cm}
\setlength{\jot}{10pt}
\titlespacing*{\section}{0pt}{4pt}{4pt}
\titlespacing*{\subsection}{0pt}{15pt}{1pt}

\fancyfoot{}
\fancyfoot[LO,RE]{\vspace{-7.1pt}}
\fancyfoot[CO]{\vspace{-7.1pt}\hspace{13.2cm}}
\fancyfoot[CE]{\vspace{-7.2pt}\hspace{-14.2cm}}
\fancyfoot[RO]{\footnotesize{{\hspace{2pt}\thepage}}}
\fancyfoot[LE]{\footnotesize{{\thepage\hspace{3.45cm}}}}
\fancyhead{}
\renewcommand{\headrulewidth}{0pt} 
\renewcommand{\footrulewidth}{0pt}
\setlength{\arrayrulewidth}{1pt}
\setlength{\columnsep}{6.5mm}
\setlength\bibsep{1pt}

\makeatletter 
\newlength{\figrulesep} 
\setlength{\figrulesep}{0.5\textfloatsep} 

\newcommand{\topfigrule}{\vspace*{-1pt}%
\noindent{\color{black}\rule[-\figrulesep]{\columnwidth}{1.5pt}} }

\newcommand{\botfigrule}{\vspace*{-2pt}%
\noindent{\color{black}\rule[\figrulesep]{\columnwidth}{1.5pt}} }

\newcommand{\dblfigrule}{\vspace*{-1pt}%
\noindent{\color{black}\rule[-\figrulesep]{\textwidth}{1.5pt}} }

\makeatother

\twocolumn[
\noindent\LARGE{\textbf{Application of pretrained universal machine-learning interatomic potential for physicochemical simulation of liquid electrolytes in Li-ion battery}}
\vspace{0.3cm} \\

\noindent\large{Suyeon Ju,$^{\ddag}$\textit{$^{a}$} Jinmu You,$^{\ddag}$\textit{$^{a}$} Gijin Kim,\textit{$^{a}$} Yutack Park,\textit{$^{a}$} Hyungmin An,\textit{$^{a}$} and Seungwu Han\textit{$^{{\ast}ab}$}} \\

\noindent\normalsize{Achieving higher operational voltages, faster charging, and broader temperature ranges for Li-ion batteries necessitates advancements in electrolyte engineering. However, the complexity of optimizing combinations of solvents, salts, and additives has limited the effectiveness of both experimental and computational screening methods for liquid electrolytes. Recently, pretrained universal machine-learning interatomic potentials (MLIPs) have emerged as promising tools for computational exploration of complex chemical spaces with high accuracy and efficiency.
In this study, we evaluated the performance of the state-of-the-art equivariant pretrained MLIP, SevenNet-0, in predicting key properties of liquid electrolytes, including solvation behavior, density, and ion transport. To assess its suitability for extensive material screening, we considered a dataset comprising 20 solvents.
Although SevenNet-0 was predominantly trained on inorganic compounds, its predictions for the properties of liquid electrolytes showed good agreement with experimental and {\it ab initio} data. However, systematic errors were identified, particularly in the predicted density of liquid electrolytes. To address this limitation, we fine-tuned SevenNet-0, achieving improved accuracy at a significantly reduced computational cost compared to developing bespoke models.
Analysis of the training set suggested that the model achieved its accuracy by generalizing across the chemical space rather than memorizing specific configurations. This work highlights the potential of SevenNet-0 as a powerful tool for future engineering of liquid electrolyte systems.
}
 \vspace{0.6cm}
  ]

\renewcommand*\rmdefault{bch}\normalfont\upshape
\rmfamily
\section*{}
\vspace{-1cm}

\footnotetext{\textit{$^{a}$~Department of Materials Science and Engineering and Research Institute of Advanced Materials, Seoul National University, Seoul 08826, Korea. E-mail: hansw@snu.ac.kr}}
\footnotetext{\textit{$^{b}$~Korea Institute for Advanced Study, Seoul 02455, Korea. }}
\footnotetext{\ddag~These authors contributed equally to this work.}

\section{Introduction}
Li-ion batteries (LIBs) have revolutionized modern technology by powering a wide range of devices, from mobile phones to electric vehicles.\cite{LIB_rev_1, LIB_rev_2} Among the various components comprising LIBs, the liquid electrolytes play a crucial role in facilitating ion transport between the anode and cathode, enabling the charging and discharging cycles.\cite{EL_1, EL_2, EL_3, EL_4, EL_5} Commercial formulations often incorporate lithium hexafluorophosphate (\ce{LiPF6}) as the Li salt,\cite{LiPF6} while ethylene carbonate (EC)-based solvents are established as the industry standard due to their ability to form a robust solid electrolyte interphase (SEI) on graphitic anodes.\cite{EC_gph} When mixed with linear carbonates such as dimethyl carbonate (DMC), ethyl methyl carbonate (EMC), and diethyl carbonate (DEC), these electrolytes offer the complementary advantages of high salt dissociation from cyclic carbonates with high permittivity, alongside the enhanced ion mobility and reduced viscosity provided by linear carbonates.\cite{KangXu_density, DMC} The introduction of vinylene carbonate (VC) as an additive has further enhanced SEI stability, with its highly reactive carbon-carbon double bond that promotes the formation of polymeric species.\cite{KangXu_density}

While the current recipes of liquid electrolytes satisfy various requirements, there is still a pressing demand for further optimizing the liquid electrolyte for LIBs to achieve enhanced energy density, safety, cycle life, and performance across various temperatures.\cite{EV_1, EV_2, EV_3} For example, to increase the operation voltage, it is necessary to lower the HOMO (Highest Occupied Molecular Orbital) level of the electrolyte to prevent degradation, or introduce additives to form a stable cathode-electrolyte interphase (CEI).\cite{HighV_1, HighV_2} On the other hand, incorporating bulky anions can increase the charging speed by elevating the Li-ion transference number,\cite{FastCharge_1, FastCharge_2} although this results in lower ionic conductivity by retarding Li-ion movements.\cite{Bulky_anions} Lastly, commercial EC-based electrolytes are vulnerable in low-temperature environments, where the viscosity increases significantly and solidification occurs, lowering the ionic conductivity.\cite{LowT_1} In addition, due to the sluggish desolvation of Li ions, charge transfer between the anode and electrolyte is hindered.\cite{LowT_2} Switching to ether-based electrolytes\cite{Ether} or utilizing (localized) high-concentration electrolytes\cite{LE_LHCE} may resolve these problems.

In the above-mentioned cases, selecting optimal formulations for liquid electrolytes often requires a careful balance between correlated material properties, such as viscosity and solvation, which calls for testing of various materials and their combinations. Considering the vast space of organic molecules and the challenges in measuring physicochemical properties experimentally, atomistic simulations have become highly useful for material screening and understanding variations in properties at the atomic level.\cite{Comp_1, Comp_2, Comp_3} In particular, molecular dynamics (MD) simulations using classical potentials or density functional theory (DFT) calculations, have been instrumental in investigating solvation structures and physicochemical properties (i.e., diffusivity and viscosity) of electrolytes,\cite{AIMD_EC_PC, AIMD_EC_EMC, AIMD_EC_cation} as well as interfacial reactions between electrodes and electrolytes.\cite{Inter_1, Inter_2, Inter_3} 

However, theoretical studies based on DFT and classical potentials face challenges in computational cost and transferability, respectively. For example, the high computational cost of DFT limits simulation size and time to a few hundred atoms and tens of picoseconds.\cite{AIMD_EC_EMC, AIMD_EC_cation} This limitation raises concerns that the simulation may not reach equilibration within the DFT time scale.\cite{ImportanceofEquil} For instance, the residence time of Li and solvent molecules can extend up to a few tens of nanoseconds,\cite{ImportanceofEquil,10.1021/acs.jpcb.3c07999} a time scale  DFT cannot practically achieve. On the other hand, while classical potentials allow simulations of tens of thousands of atoms over hundreds of nanoseconds, they sacrifice transferability and general accuracy by fitting model parameters to DFT results or experimental data specific to particular systems. For example, the charges in OPLS-AA (Optimized Potentials for Liquid Simulations-All Atom) were scaled by 80\% to fit to the experimental Li diffusivities in EC electrolytes.\cite{doi:10.1021/acs.jctc.6b00824, doi:10.1021/acsenergylett.9b02118} However, this approach deteriorated the diffusivities of both the Li ion and the PF$_6^-$ anion in PC solvent.\cite{doi:10.1021/acs.jctc.6b00824} Similar trade-offs on different properties have been reported in other classical potentials, such as in TraPPE (Transferable Potentials for Phase Equilibria) force fields,\cite{TraPPE} which accurately predicted densities but showed more than 20\% error in relative permittivities. Beyond these Class I force fields, more advanced force fields, such as Class II\cite{Class-II} incorporating bond and angle anharmonicity, and APPLE\&P\cite{APPLEP_2009} (Atomistic Polarizable Potential for Liquids, Electrolytes, and Polymers, Class-III), which is many-body polarizable, achieve higher accuracy through additional parameters, but they demand careful parameter tuning and still suffer from limited transferability.

Over the past decade, data-driven machine learning interatomic potentials (MLIPs) have gained significant attention in materials simulation by extending both the length and time scales to those of classical potentials while maintaining accuracy close to that of DFT.\cite{doi:10.1021/acs.chemrev.1c00022, doi:10.1021/acs.chemrev.0c01111, doi:10.1080/27660400.2023.2269948, doi:10.1021/jacs.3c06210, doi:10.1021/acscatal.3c04964} Therefore, employing MLIPs in the simulation of liquid electrolytes is poised to overcome the difficulties in DFT or classical force fields mentioned above.
However, there are significant challenges in developing MLIPs for liquid electrolytes, particularly with traditional application-specific, bespoke-style MLIPs. First, generating training sets with DFT-based MD simulations incurs high computational costs. This is because adequate sampling of all possible configurations, including different molecular conformers and achieving ergodicity, requires long-term simulations.  
Second, in order to computationally identify optimal formulations from material screening, it is necessary to develop MLIPs that can be applied to a wide range of organic molecules. This in turn requires the creation of a comprehensive training set incorporating various combinations of solvents and salt pairs and a careful sampling of both intramolecular and intermolecular interactions among different chemical moieties. Thus, previous studies have relied on many cycles of iterative learning to generate such training sets,\cite{QRNN, Csanyi} incorporating various strategies to improve the precision of intermolecular interactions. Consequently, most studies using MLIPs have been limited to investigating specific solvent-salt systems, such as glyme-based electrolytes\cite{glyme} and carbonate electrolytes,\cite{QRNN, Degradation} or solvent-only systems consisting of mixtures of EC and DMC\cite{Csanyi}. In another example using the graph neural network interatomic potential (GNN-IP), ref~\citenum{DeepEE} analyzed the Li transport mechanisms in deep eutectic electrolytes and lithium bis(trifluoromethanesulfonyl)imide (\ce{LiTFSI}). It is notable that ref~\citenum{BAMBOO} has extended the chemical space by incorporating ester materials and fluorine doping at various sites in carbonates. However, for untrained fluorine-doped systems, the density error reached a maximum of 21\%.

Recently, pretrained general-purpose GNN-IPs such as M3GNet,\cite{M3GNet} CHGNet,\cite{CHGNet} PFP,\cite{MATLANTIS} GNoME,\cite{GNoME} MACE-MP-0,\cite{MACE-MP-0} SevenNet-0,\cite{SevenNet} MatterSim,\cite{MatterSim} eqV2 M,\cite{eqV2M} and ORB\cite{Orb} have emerged, providing generalizability across diverse chemical spaces. It has been also shown that fine-tuning the pretrained model can achieve the precision of bespoke models at a small cost.\cite{PerformanceAssessment, Sublimation, SystematicSoftening} The generalizability of these models largely stems from the architecture of GNN-IPs, such as NequIP\cite{NequIP} and MACE,\cite{MACE} which automatically extracts important features from deep learning. In addition, atomic species are embedded with learnable parameters, allowing the model to learn chemical similarities between elements.\cite{Roadmap} This enables the pretrained model to capture general trends in chemical bonding. 

Most of current pretrained models were trained using inorganic materials databases such as the Materials Project,\cite{MaterialsProject} Alexandria,\cite{Alexandria} and OMat24.\cite{eqV2M}
Nevertheless, MACE-MP-0 demonstrated reasonable accuracy and stability in simulating liquid electrolytes when applied to the EC/EMC \ce{LiPF6} electrolyte and a complete battery system.\cite{MACE-MP-0} In another example, ref~\citenum{MACE_Csanyi} compared densities and diffusivities of 3:7 EC:EMC solvents between the bespoke MACE and MACE-MP-0 models and found resonable agreements. However, for pretrained models extensively trained on inorganic compounds, liquid electrolytes fall into strongly out-of-distribution domains. First, as will be detailed in this work, relevant structural motifs of organic molecules were not fully sampled in the dataset. Furthermore, most pretrained models do not explicitly account for long-range electrostatic interactions, making it unclear how well they can describe solvation structures and dielectric properties, which are primarily governed by Coulomb interactions. Therefore, a systematic analysis on the general accuracy of current pretrained models in liquid electrolytes is on demand.

In this work, motivated by the above discussions, we systematically investigate the accuracy of SevenNet-0\cite{SevenNet-0} (simply SevenNet or 7net henceforth) in applications involving liquid electrolytes. SevenNet has achieved high performance in the Matbench Discovery benchmark, which assesses the performance of pretrained universal force fields on inorganic crystal discovery.\cite{Matbench} To evaluate whether the pretrained model is suitable for extensive material screening, we tested its performance across a diverse range of electrolytes. Key properties such as densities, solvation shell structures, and diffusivities were compared with experimental data or \textit{ab initio} MD (AIMD) results. While the overall agreement with reference data was good, SevenNet exhibited force softening and overestimated solvent density.
To address this, we also fine-tuned SevenNet, which significantly improved the model's accuracy. This work highlights the potential of SevenNet for future  engineering of liquid electrolyte systems.

\section{Results and discussions}
\begin{figure}
\centering
  \includegraphics[width=8.3cm]{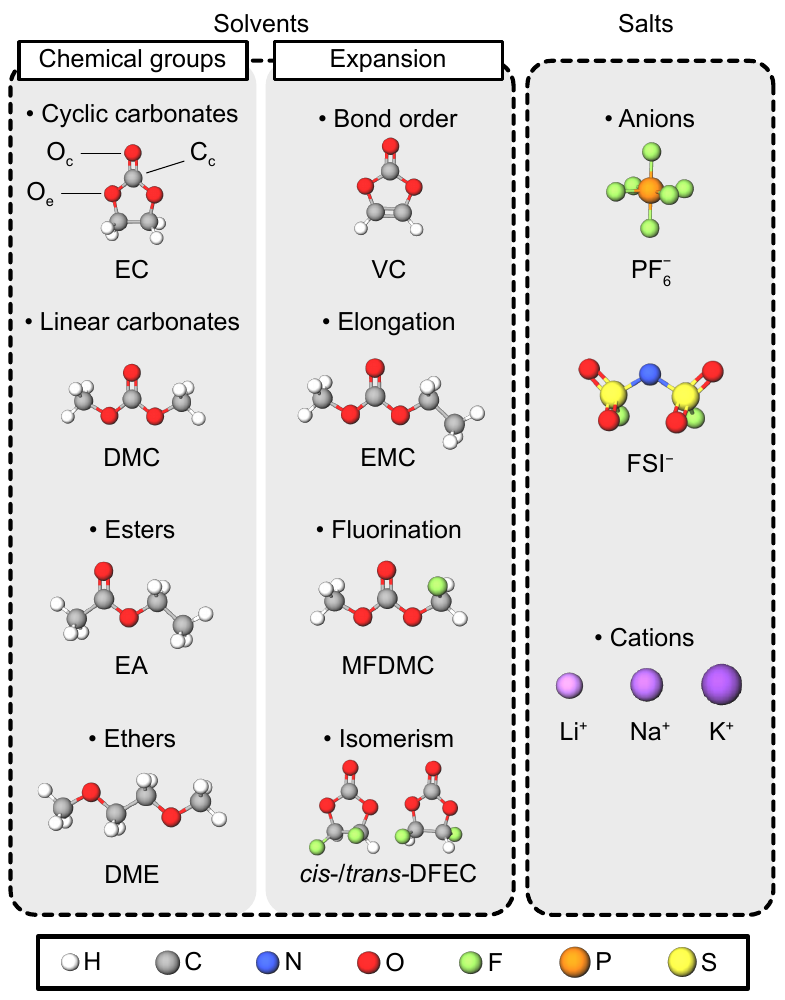}
  \caption{Schematic description of the solvent molecules and ions investigated in this study. The solvents are categorized by chemical groups (carbonate, ester, and ether) and molecular structures (cyclic or linear) (left). Variations in the solvent set include changes in bond order, molecular elongation, partial fluorination, and isomerism (middle). Examples of the carbonyl oxygen (\ce{O_c}), carbonyl carbon (\ce{C_c}), and ethereal oxygen (\ce{O_e}) are indicated on the EC molecule. The anions and cations used for salts in the electrolyte systems are shown in the right. For corresponding IUPAC names and formulas, refer to Table S1.\dag}
  \label{fgr:structures}
\end{figure}

To test SevenNet with various types of molecules employed in liquid electrolytes, we considered a total of 20 solvents and 2 salts, as listed in Table S1\dag\ and schematically summarized in Fig.~\ref{fgr:structures}. The test solvents were selected to encompass a broad range of liquid electrolytes used in commercial batteries or in advanced battery research. They represent four major chemical groups: cyclic carbonates, linear carbonates, ethers, and esters. As base molecules, we used EC for cyclic carbonates, DMC for linear carbonates, dimethoxyethane (DME) for ethers, and ethyl acetate (EA) for esters, as shown on the left of Fig.~\ref{fgr:structures}. To expand the chemical space, we varied the molecular structures by altering the bond order, elongating carbon chains, incorporating fluorine atoms, and exploring \textit{cis–trans} isomerism, as depicted in the center of Fig.~\ref{fgr:structures}. For salts, \ce{LiPF6} and lithium bis(fluorosulfonyl)imide (\ce{LiFSI}) were considered (see the right of Fig.~\ref{fgr:structures}).
For cations, we focused primarily on Li ions (Li$^+$), although sodium (Na$^+$) and potassium 
(K$^+$) ions were also included in the solvation shell analysis.

The full simulation of liquid electrolytes involves various types of bonding/nonbonding, intra-/intermolecular interactions. To systematically assess the accuracy of SevenNet, we apply the model to progressively more complex systems in the following subsections, starting from single molecules, moving to pure solvents, and finally to full electrolytes.

\subsection{Single solvent molecules}
\begin{figure*}
\centering
  \includegraphics[height=10cm]{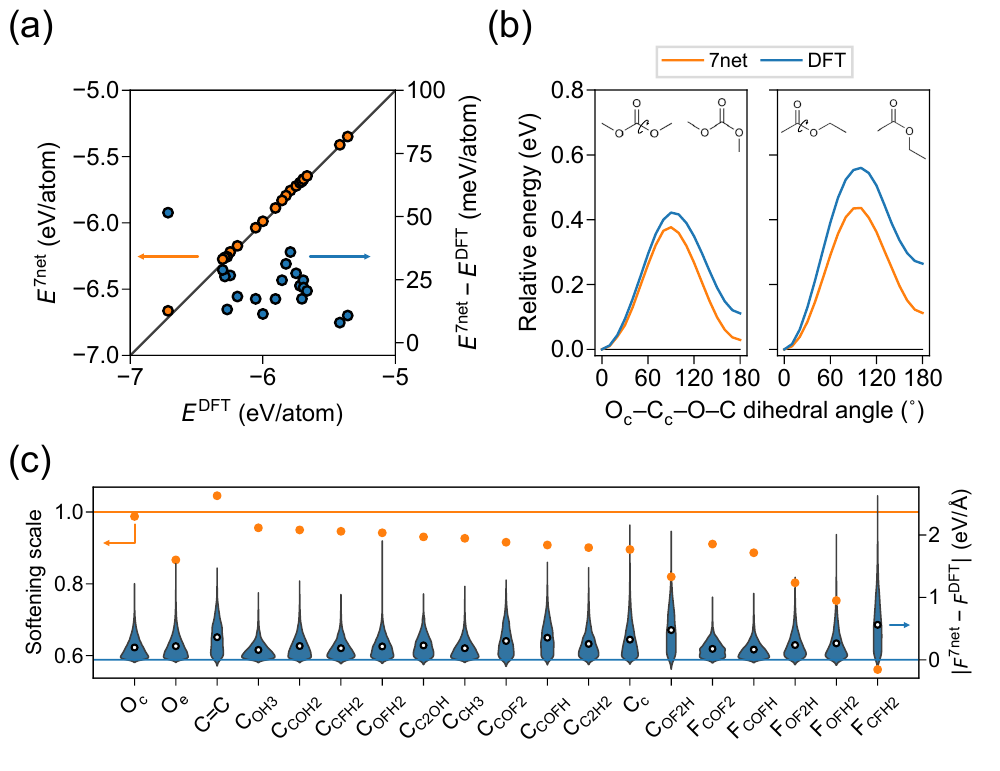}
  \caption{(a) Comparison of per-atom energies (left axis) and corresponding errors (right axis) between DFT and SevenNet predictions for 20 single solvent molecules. (b) Relative energy profiles for DMC (left) and EA (right) molecules as a function of \ce{O_c-C_c-O-C} dihedral angle. See the inset for the schematic images. (c) Force softening scales for each atom type (left axis) and the distribution of absolute errors in force components (right axis) obtained from single-molecule SevenNet MD trajectories. MAEs of the force components are indicated by white-filled circles. }
  
  \label{fgr:Intramolecular}
\end{figure*}

For the simulation of organic systems, an accurate description of a single molecule is a basic requirement. We compared the single-molecule energies and structures of SevenNet and DFT for 20 solvent molecules. 
The molecule was initially placed in a cubic box with periodic boundary conditions, where the length of the box was set to 10 Å plus the maximum molecular length along each axis. In the DFT calculations, a dipole correction was applied in all directions to remove spurious dipole-dipole interactions between periodic images. Structural relaxation was performed until the magnitude of the atomic forces was reduced to within 0.02 eV/Å. In Fig.~\ref{fgr:Intramolecular}a, we compare the per-atom energies obtained by SevenNet and DFT relaxation. The mean absolute error (MAE) is 23 meV/atom, where the largest deviation of 51 meV/atom found with the VC molecule. These relatively small errors do not alter the relative energy ordering of the single molecules, offering basic evidence of accuracy of SevenNet. We also compared the bond lengths and angles of relaxed molecular structures to verify that SevenNet produced molecular geometries correctly. The MAEs for bond lengths and angles are 0.005 Å and 0.7°, respectively, indicating highly accurate structural predictions.

Linear solvents exhibit \textit{cis–trans} conformers, and structural differences affect dipole moments,\cite{Dipole_1, Dipole_2} influencing the participation of solvent in Li-ion solvation.\cite{WhyDMC} 
We select two molecules, DMC and EA, since linear carbonates and esters generally exhibit larger energy barriers between conformers compared to other types of molecules.
In Fig.~\ref{fgr:Intramolecular}b, by performing dihedral-angle-constrained ionic relaxations, we scan a total of 180° in 10° intervals for the \ce{O_c-C_c-O-C} dihedral angle in DMC and EA (see the inset). It is seen that SevenNet underestimates the torsion barrier by 0.04 eV and 0.12 eV for DMC and EA, respectively. This underestimation of barriers may be related to the softening of potential energy surface (PES) in pretrained models that were primarily trained on low-energy structures.\cite{SystematicSoftening} 

To investigate the accuracy of atomic forces, we conducted a 0.5-ns MD simulation of a single molecule in an NVT ensemble using SevenNet. A Nosé--Hoover thermostat\cite{NH_thermo} and a timestep of 0.5 fs were employed. The temperature was set to 600 K to sample high-energy structures, and 500 snapshots were extracted at intervals of 1 ps. Subsequently, single-point DFT calculations were performed on these snapshots.
The softening scales and the corresponding absolute error distributions of the force components are shown in Fig.~\ref{fgr:Intramolecular}c.
The softening scale is defined as the slope of the linear function fitted to the force parity plot. The ideal value is one, and those below one indicate that the forces predicted by SevenNet were systematically smaller than those calculated by DFT.
To examine whether force error and softening depend on specific atom types, atoms were classified into 19 categories (Fig.~S1\dag). To be specific, the oxygen atoms were divided into carbonyl (\ce{O_c}) and ethereal (\ce{O_e}) oxygens (see Fig.~\ref{fgr:structures}). The carbon atoms were categorized based on their bonding environments; for example, \ce{C_{COFH}} represents a carbon atom bonded to one carbon, one oxygen, one fluorine, and one hydrogen atom. Double-bonded carbons in VC were classified separately. Fluorine atoms were labeled according to the type of carbon atom to which they are bonded; for instance, \ce{F_{COFH}} refers to fluorine atoms bonded to a \ce{C_{COFH}} carbon atom.
In Fig.~\ref{fgr:Intramolecular}c, most atom types exhibit some degree of force softening; however, the extent of softening varies among atom types. Notably, pronounced softening is observed for fluorine atoms in \ce{F_{OFH2}} and \ce{F_{CFH2}} local structures. As will be discussed below, this significant level of softening may stem from the limited representation of these chemical moieties in the training set.

\subsection{Pure solvents}
Next, we simulated pure solvents composed of a single type of organic molecule to obtain their theoretical densities. Liquid density plays a critical role in determining the physicochemical properties of electrolytes. For example, a density decrease of just 0.1 g/cm$^{3}$ can result in a twofold increase in diffusivity for acetonitrile at 298 K.\cite{acetonitrile}
The initial configurations for the liquid simulations were generated using MolView\cite{Molview} and PACKMOL.\cite{PACKMOL} The number of molecules was chosen so that the total number of atoms was closest to 1000. To determine the length of the cubic simulation box without relying on experimental data, we first obtained $V_0$ by adding the van der Waals volumes\cite{10.1021/jp8111556} of all atoms in the simulation box. The initial box length was then set to  $1.1 \times V_0^{1/3}$. (The actual numbers of molecules and the sizes of the simulation boxes for each simulation are summarized in Table S2.\dag) For propylene carbonate (PC) and fluoroethylene carbonate (FEC), two chiral conformers were considered in equal proportions. For the \textit{cis–trans} conformers of linear carbonates and the \textit{syn}/\textit{anti} conformers of esters, the Boltzmann distribution, with potential energies calculated using SevenNet, was used to determine the initial ratio (see Tables S3 and S4\dag).

The initial structures were relaxed under loose conditions ($\max_{i}{|\mathbf{F}_{i}|}$ < 2.0 eV/Å), followed by MD simulations for 1 ns in the NPT ensemble with a timestep of 2 fs at the target temperature and a pressure of 1 atm. The Nosé--Hoover thermostat and barostat\cite{NH_baro} were applied as implemented in the \texttt{LAMMPS} package.
To ensure a fair comparison with experimental data, the simulation temperatures were matched to the experimental conditions. Specifically, all simulations were conducted at 298 K, except for EC (313 K), FEC (313 K), DME (293 K), and 1,2-diethoxyethane (DEE) (293 K).
To maintain stable simulations and prevent large positional fluctuations with a timestep of 2 fs, the atomic mass of tritium ($^3\text{H}$, 3 a.u.) was assigned to hydrogen atoms.
To determine the equilibrium density of the system, an additional 0.4-ns simulation was performed in the NPT ensemble with a reduced timestep of 1 fs. The density was calculated by averaging the instantaneous density values (recorded every 10 fs) during the last 0.2 ns of the simulation.
Throughout all equilibration and production runs, no spurious reactions were observed, confirming the stability of the simulations.

\begin{figure}
\centering
  \includegraphics[width=6cm]{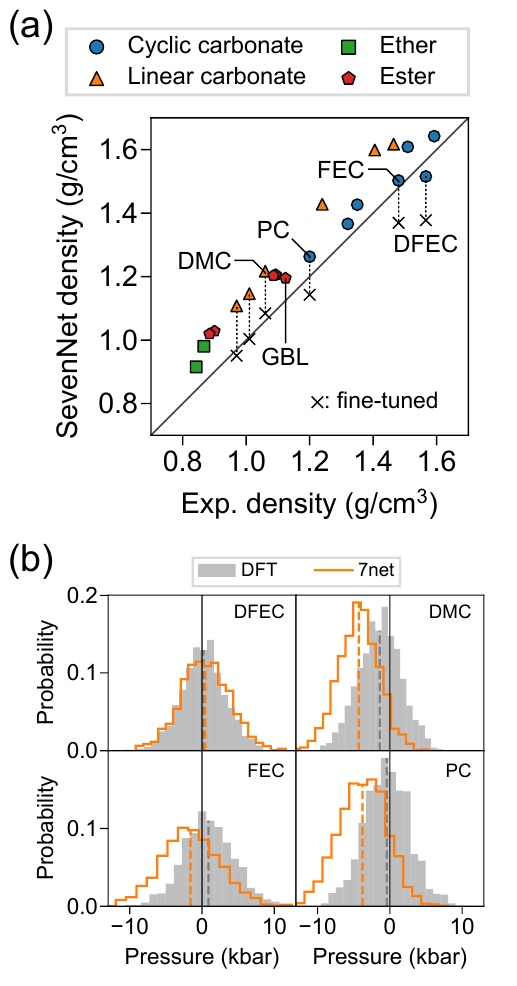}
  \caption{(a) Parity plot comparing pure solvent densities calculated using SevenNet to experimental values, categorized by chemical groups. MD simulations were conducted at 298 K, except for EC (313 K), FEC (313 K), DME (293 K), and DEE (293 K). The blue circles, orange triangles, green squares, and red pentagons represent cyclic carbonates, linear carbonates, ether, and ester, respectively. The black `x' symbols indicate the values obtained with fine-tuned SevenNet.
  (b) Pressure distributions for DFEC, FEC, DMC, and PC from individual SevenNet-MD and AIMD simulations conducted in the NVT ensemble over 15 ps at experimental densities. Temperatures are set to match the experimental values in Table S5.\dag\ Vertical dashed lines indicate the mean values of each distribution.
  }
  \label{fgr:pure_density}
\end{figure}

Fig.~\ref{fgr:pure_density}a presents computed liquid densities. Since reference densities obtained with DFT are scarce, we compared with experimental data sourced from refs~\citenum{KangXu_density, PCVC_density, FEC_density, BAMBOO, DME_density, DEE_density, EA_density, EP_density, GBL_density}.
(See Table S5\dag\ for the actual values and errors.) 
To identify systematic trends, we classified the solvents into four chemical groups; cyclic carbonates, linear carbonates, ethers, and esters. 
Fig.~\ref{fgr:pure_density} shows that the densities computed using SevenNet generally overestimates compared to experimental values. However, the degree of overestimation is not random but consistent within chemical groups. To be specific, cyclic solvents, including cyclic carbonates and $\gamma$-Butyrolactone (GBL), exhibit smaller deviations, ranging $-$3\%$\sim$7\%, while linear solvents show higher overestimation by 9\%$\sim$15\%. 

To clarify whether the discrepancy in predicted densities is primarily due to limitations of the SevenNet model or inherent inaccuracies in the DFT functional (PBE-D3), we examined the pressure distributions (average of diagonal components of the virial stress) by SevenNet and DFT for FEC, difluoroethylene carbonate (DFEC), DMC, and PC liquids, when the volumes are adjusted to match with the experimental densities. The temperatures were set to 298 K (DEFC, DMC, and PC) or 313 K (FEC). For the feasibility with DFT calculations, we chose a smaller simulation cell containing 30 solvent molecules and the 
initial configurations  (with equal chirality for PC and FEC) were equilibrated under SevenNet for 1 ns.  
From these equilibrated structures, we ran 20-ps production simulations in NVT ensembles independently with either SevenNet or DFT with a 1-fs timestep, recording the pressure every 10 fs during the final 15 ps to gather 1500 data points.

Fig.~\ref{fgr:pure_density}b shows the pressure distributions of both DFT and SevenNet for each solvent, with vertical dashed lines marking the mean values of each distribution. For DFT, the average pressure remains close to zero in all solvents, suggesting that the experimental volume is close to the equilibrium density predicted by PBE-D3. The maximum absolute pressure is 1.4 kbar in DMC. Assuming a bulk modulus of roughly 1.6 GPa at 298 K,\cite{DMC_bulk_modulus} the corresponding density error of DFT is on the order of 9\%, although a more rigorous convergence study exploring larger simulation cells and longer time scales, would be needed for quantitative accuracy. A previous work demonstrated that bespoke MLIPs can predict densities of EMC-rich solvent mixtures to within about 5\% of experiment at the PBE-D3 level, using a relatively small cutoff radius of 10 Å for the D3 term.\cite{Csanyi}

In contrast, SevenNet predicts more negative pressures for DMC and PC in Fig.~\ref{fgr:pure_density}b, indicating a significant compressive stress that drives the system toward higher density. This shift is correlated with the degree of density overestimation observed for PC and DMC when using SevenNet (see Fig.~\ref{fgr:pure_density}a). Further single-point DFT calculations on SevenNet-generated snapshots also confirmed that these pressure discrepancies arise from inaccuracies in the SevenNet model rather than from fundamental errors in the DFT reference (Fig.~S2\dag); the trends in both mean error and average pressure aligned closely. Consequently, while slight errors in the DFT reference cannot be completely ruled out, we conclude that the dominant source of density overestimation in SevenNet is an imperfect learning of stress. As discussed in the following section, we found that fine-tuned SevenNet reduced this discrepancy, suggesting that improved training strategies can bring the predicted pressures and densities closer to the DFT and also experimental values.

In Fig.~S3 and S4,\dag\ we calculated potential energy curves for EC and DMC dimers across eight types of interactions (H--H, $\mathrm{O_c}$--H, $\mathrm{O_c}$--$\mathrm{O_c}$, $\mathrm{O_c}$--$\mathrm{O_e}$, $\mathrm{O_e}$--H, $\mathrm{O_e}$--$\mathrm{O_e}$, orthogonal, and planar orientations). It is found that SevenNet has a deeper potential well near equilibrium, with some equilibrium distances being shorter than those from DFT. 
Thus, the pressure deviations in the above can be attributed to the stronger intermolecular bonding in SevenNet compared to DFT.

We compared the performance of SevenNet in predicting liquid densities with other computational methods, namely QRNN,\cite{QRNN} OPLS4,\cite{QRNN} and BAMBOO\cite{BAMBOO} (Fig.~S5a\dag). QRNN and OPLS4 are quantitatively more accurate, showing errors of $-$5\%$\sim$2\% and $-$2\%$\sim$4\%, respectively. We note that a portion of the QRNN training set was generated at high pressures and another portion with $\pm$20\% scaling in intermolecular distances, implying that QRNN may have learned the equilibrium volume from these training sets. The BAMBOO model generally provided density estimates closer to experimental data than SevenNet, with the exception of the fluorinated linear carbonate group. A density alignment method was employed for BAMBOO, wherein the model was trained to match experimental densities for systems such as EC, PC, FEC, DEC, DMC, and EA, which are plotted together in Fig.~S5a.\dag\ While this alignment method enabled accurate predictions for in-domain systems, it limited generalizability to less similar systems, as evidenced by the bis(fluoromethyl) carbonate (DFDMC) case, where the model overestimates density by 20\%. In contrast, SevenNet exhibited moderate and regularized accuracy across all solvent systems.

\subsection{Full electrolytes}
In this subsection, we applied SevenNet to simulate full electrolytes composed of solvents and ion salts. First, we investigated the solvation shell structures for dilute electrolytes. Next, at a conventional concentration of Li salt, we analyzed the effect of solvent type on Li solvation, with a focus on salt dissociation. Finally, diffusion coefficients were obtained and compared with experimental values.

\subsubsection{Solvation shell structures in dilute solutions}
The solvation structure of ions plays a critical role in electrolyte systems, influencing both ion dissociation and the stability of electrode interfaces, which directly impacts ionic conductivity and battery cycle life. In ref~\citenum{AIMD_EC_EMC}, solvation shell structures around Li ions were investigated in detail by AIMD. In this subsection, we benchmark SevenNet against these reference results by adopting the same simulation protocol. In detail, the initial structure consisted of 63 EC (42 EMC) molecules and one \ce{LiPF6} salt, corresponding to dilute conditions (0.2 M). Experimental densities of pure EC (1.32 g/cm$^3$ at 313 K) and EMC (1.01 g/cm$^3$ at 298 K) were used,\cite{KangXu_density} yielding cubic-cell lengths of 19.28 Å for EC and 19.52 Å for EMC. To obtain various solvation types, ten independent simulations were performed: five starting with a dissociated ion pair and the other five with an associated pair. Each structure was equilibrated for 7.5 ps at 330 K in the NVT ensemble, followed by a 30-ps production run with a timestep of 0.5 fs.  Here, the dispersion interaction was excluded like ref~\citenum{AIMD_EC_EMC}.  Snapshots were saved every 5 fs, generating 6001 snapshots for each production run. A total of 60 010 snapshots were collected for each solvent.

In liquid electrolytes, Li ions undergo various types of solvation environments since there can be multiple coordinating oxygens in a solvent molecule, and anion can also be introduced in the solvation shell. Each solvation shell type has distinct structural features, like  radial and angular distributions related to \ce{Li-O}, as studied by AIMD.~\cite{AIMD_EC_EMC} 
Following the reference, we classified the collected snapshots into five types of solvation shell: \ce{Li(EC_c)4}, \ce{Li(EC_c)3(PF6)}, \ce{Li(EMC_c)3(PF6)}, \ce{Li(EMC_c)3(EMC_e)}, and \ce{Li(EMC_e)2(PF6)}. 
Such classification was based on the composition of the first solvation shell of a Li ion and the coordinating oxygen type. For example, if a Li ion is coordinated with four EMC molecules--three via \ce{O_c} and one via \ce{O_e}--then it is designated as \ce{Li(EMC_c)_3(EMC_e)}. When both \ce{O_c} and \ce{O_e} of a single solvent molecule simultaneously coordinated a Li ion, the coordinating oxygen type was designated as \ce{O_c}, where coordination by two \ce{O_e} atoms was classified as \ce{O_e}-type coordination.

The radial distribution function (RDF) $g(r)$ and the coordination number (CN) of an atom type B around an atom type A is defined as follows:\cite{10.1021/acs.chemrev.1c00904}
\begin{align}
\label{eqn:RDF}
g(r) &= \frac{1}{\rho_{\mathrm{B}}} \frac{1}{4\pi r^2} \frac{dN(r)}{dr} \\
\label{eqn:CN}
\mathrm{CN} &= N(r_\mathrm{cut})
\end{align}
where $N(r)$ is the average number of particle B within a sphere centered on particle A with a radius of $r$, and $\rho_{\mathrm{B}}$ is the number density of B atoms. The cutoff threshold $r_\mathrm{cut}$ of \ce{Li-O} was set to 2.6 Å and 4.2 Å for \ce{Li-P}, which were obtained from the first minimum of RDF.

\begin{figure}
\centering
  \includegraphics[width=8.3cm]{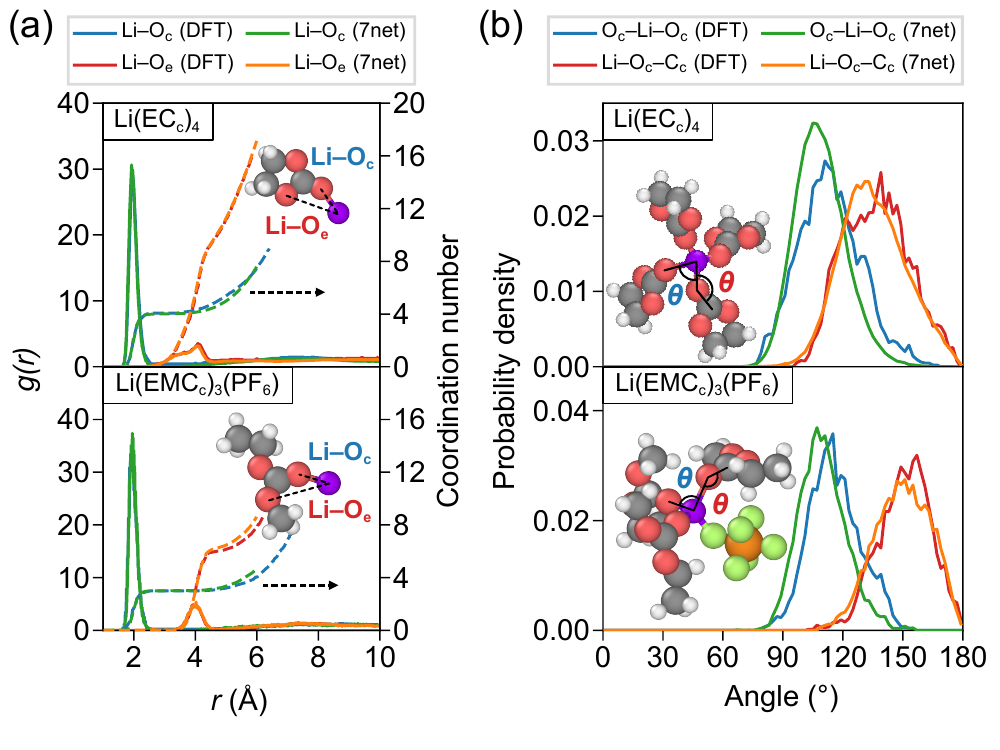}
  \caption{Solvation structures of \ce{Li(EC_c)_4} and \ce{Li(EMC_c)_3(PF_6)} solvation types obtained by SevenNet, compared with DFT results.\cite{AIMD_EC_EMC} (a) RDFs (solid lines) of \ce{Li-O_c} and \ce{Li-O_e}, along with their CNs (dashed lines) for each solvation type. Examples of \ce{Li-O_c} and \ce{Li-O_e} distances are illustrated in the inset. (b) Angular distributions of \ce{O_c-Li-O_c} and \ce{Li-O_c-C_c} angles for each solvation type. Examples of angles are illustrated in the inset.
  }
  \label{fgr:shell}
\end{figure}

The bond length distributions of Li and oxygen of two representative solvation types, \ce{Li(EC_c)4} and \ce{Li(EMC_c)3(PF6)}, are presented in Fig.~\ref{fgr:shell}a. (The corresponding results for other solvation types are shown in Fig.~S6.\dag) Overall, the distributions of \ce{Li-O_c} and \ce{Li-O_e} distances and CNs (right axis) show good agreements with DFT results. In experiments, slightly longer bond lengths of 2.04--2.08 Å were reported in concentrated solutions of \ce{LiPF6} salt in PC and DMC.\cite{Li-O_PC_exp,Li-O_DMC_exp}

The distributions of \ce{O_c-Li-O_c} and \ce{Li-O_c-C_c} angles in Fig.~\ref{fgr:shell}b also agree well with DFT. The \ce{O_c-Li-O_c} angle peaked at approximately 107°, minimizing steric hindrance within the near-tetrahedral coordination environment. This angle remains consistent regardless of the presence of PF$_6^-$ in the first solvation shell.
Meanwhile, the \ce{Li-O_c-C_c} angle deviates from 180°, likely due to the partial negative charges on the two \ce{O_e} atoms bonded to \ce{C_c}, which attract positively charged Li ions. This angle varies depending on the solvent: 132° in EC and ranging from 149° to 156° in EMC. This difference reflects the steric hindrance imposed by the bulky, linear structure of the EMC molecule.
A similar trend has been observed experimentally, with PC exhibiting a \ce{Li-O_c-C_c} angle of 138° and DMC showing an angle of 153°.\cite{Li-O_PC_exp,Li-O_DMC_exp} On the other hand, ReaxFF produced a sharper peak near 90°, suggesting a more rigid \ce{Li-O_e} interaction,\cite{AIMD_EC_EMC} highlighting limitations in ReaxFF.

We extended our study to examine other alkali metal ions, \ce{Na+} and \ce{K+}. By replacing \ce{Li} with \ce{Na} or \ce{K} in 63 EC + 1 \ce{LiPF6} simulations, we conducted a 7-ps equilibration MD run followed by a 25-ps production run for each cation.
The distributions of the cation-oxygen distance are shown in Fig.~S7.\dag\ The Na-O and K-O bond lengths are 2.34 and 2.74 Å, respectively, aligning closely with AIMD results of 2.35 and 2.80 Å.\cite{AIMD_EC_cation} Compared to the Li-O distribution, the first peak is broadened and reduced in intensity, indicating weaker interactions between the cation and oxygen atoms as the ionic radius of the cation increases. This weaker solvation of \ce{Na+} and \ce{K+} relative to \ce{Li+} arises from the delocalization of outer-shell electrons in larger cations, which hinders lone-pair sharing from oxygen atoms.\cite{AIMD_EC_cation}

These results demonstrate that SevenNet effectively recognizes the Li atom in the solvent as a cation and accurately captures Coulomb interactions with negatively charged oxygen atoms, despite being trained without explicit charge information.

\subsubsection{Effect of solvent type on Li solvation}
In this subsection, we investigated electrolytes with high Li salt concentrations used in commercial batteries. Depending on the solvents, the degree of ion dissociation between cations and anions varies, resulting in structures such as solvent-separated ion pairs (SSIPs), contact ion pairs (CIPs), and aggregates (AGGs), as illustrated in Fig.~S8a.\dag\ Mixing cyclic solvents with high dielectric constants and linear solvents with lower dielectric constants influences ion dissociation and the composition of the first solvation shell.\cite{Hayamizu_2017, Binary_Exp} Accurately simulating these phenomena is essential for identifying optimal electrolytes, which often balances multiple objectives, such as high ionic conductivity and robust SEI formation.\cite{KangXu_density, EL_1}

As a concrete example, we selected the EC/DMC binary solvent system with 1 mol/kg \ce{LiPF6}, varying the EC molar fraction, $x_{\mathrm{EC}} = N_{\mathrm{EC}} / (N_{\mathrm{EC}} + N_{\mathrm{DMC}})$, where $N_\alpha$ is the number of $\alpha$ molecules. This system has been widely studied to explore the competition between two different solvent types within the Li solvation shell.\cite{APPLEP_2009,AIMD_EC_DMC_conc,Binary_Exp,Binary_CFF} Recent experimental and computational studies have confirmed that EC is preferred for Li solvation over DMC, reducing cation-anion cross-correlation and enhancing ionic conductivity, particularly at low $x_{\mathrm{EC}}$ values.\cite{Binary_Exp,Binary_CFF}

All simulations were performed in the NVT ensemble at experimental densities. (Like the case with pure solvents, SevenNet overestimated the density of solvent-salt systems, as shown in Fig.~S5b\dag\ for EC/LiFSI electrolyte.)  The conformer ratio of DMC for each composition was obtained from the experiment.\cite{Binary_Exp} The temperature was set to 298 K to match the experimental conditions. The number of solvent and salt molecules was adjusted to correspond to a \ce{LiPF6} concentration of 1 mol/kg, ensuring the total number of atoms remained below 1000. Li ions were placed by dividing the simulation domain into distinct regions and randomly distributing the Li ions within these regions, as illustrated in Fig.~S9.\dag\ This approach prevented the generation of initial structures where Li ions were clustered on one side. The specific numbers of solvent and salt molecules, along with the simulation box lengths, are summarized in Table S6.\dag\
Equilibration was performed over 1.4 ns, starting with an initial 1 ns using a 2-fs timestep and a hydrogen atomic mass of 3 a.u., as described earlier, followed by an additional 0.4 ns of equilibration. A production run of 1 ns was conducted, with snapshots saved every 100 fs. The CNs for EC, DMC, and PF$_6^-$ anions were calculated and averaged across all snapshots for each simulation. To ensure statistical reliability, three independent runs were carried out, each starting from a distinct initial configuration.

\begin{figure}
\centering
  \includegraphics[width=7.5cm]{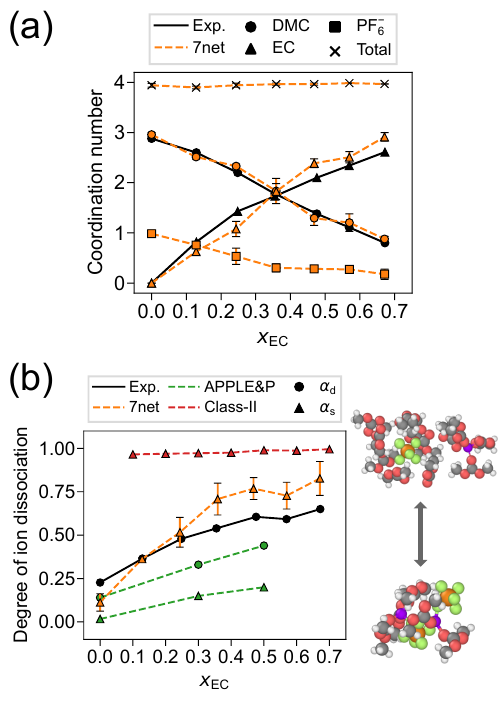}
  \caption{Variations in solvation environment in the EC/DMC binary solvent system with 1 mol/kg \ce{LiPF6}.
  (a) Average CNs of Li contributed by EC and DMC solvents and (b) the degree of ion dissociation as a function of $x_{\textrm{EC}}$. Experimental values for 1 mol/kg \ce{LiPF6}\cite{Binary_Exp} are shown alongside results from other force fields (APPLE\&P\cite{APPLEP_2009} and Class-II force field\cite{Ravikumar}) for 1 M \ce{LiPF6}. Circle markers represent the dynamic degree of dissociation $\alpha_{\textrm{d}}$, while triangle markers represent the static degree of dissociation $\alpha_{\textrm{s}}$. Error bars represent the standard deviation of 3 simulations. Representative solvation structures for fully dissociated and aggregated cases are displayed on the right side of the plot. 
  }
  \label{fgr:Solvation}
\end{figure}

Fig.~\ref{fgr:Solvation}a shows the computed compositions of the Li solvation shell as a function of $x_{\mathrm{EC}}$. The computational results align well with experimental trends\cite{Binary_Exp} and classical MD simulations (not shown).\cite{APPLEP_2009,AIMD_EC_DMC_conc} Notably, EC molecules were strongly favored in the solvation shell at low $x_{\mathrm{EC}}$. For instance, at $x_{\mathrm{EC}} =$ 0.13, the local fraction of EC in the Li solvation shell is 0.25 (SevenNet) and 0.32 (experiment), significantly higher than the bulk EC ratio of 0.13. Over the entire range of 0$ < x{_\mathrm{EC}} < $0.7, the total CN of Li remains approximately 4. A comparison of CNs obtained using different force fields is presented in Fig.~S10.\dag\

Next, we computed the degree of ion dissociation, whose definition subtly varies across the literature.\cite{APPLEP_2009, mdpi} The static degree of ion dissociation, $\alpha_{\mathrm{s}}$, is defined as the fraction of free ions,\cite{APPLEP_2009}  which in turn is defined as having no counterions in its first solvation shell, corresponding to the SSIP solvation state.
Some studies focused specifically on free cations,\cite{mdpi} which tends to yield slightly higher values of $\alpha_{\mathrm{s}}$ compared to free ions.\cite{APPLEP_2009}
On the other hand, the dynamic degree of ion dissociation, $\alpha_{\mathrm{d}}$, also referred to as the degree of uncorrelated motion, is defined as the ratio of ionic conductivity to the Nernst-Einstein conductivity.\cite{APPLEP_2009, Hayamizu_2017} While $\alpha_{\mathrm{s}}$ and $\alpha_{\mathrm{d}}$ are not directly interchangeable, they exhibit similar trends with varying solvent composition.\cite{Binary_Exp, APPLEP_2009} For instance, in the GBL/DMC system with LiFSI salts,\cite{Binary_Exp} $\alpha_{\mathrm{s}}$ and $\alpha_{\mathrm{d}}$ were found to be comparable at $x_{\mathrm{GBL}} >$ 0.4. However, as $x_{\mathrm{GBL}}$ approached zero, $\alpha_{\mathrm{s}}$ decayed more rapidly than $\alpha_{\mathrm{d}}$.
In our study, we employed the fraction of free cations, $\alpha_{\mathrm{s}}$, to represent the degree of ion dissociation, as calculating $\alpha_{\mathrm{d}}$ requires long-term MD simulations.

Fig.~\ref{fgr:Solvation}b presents computational results alongside experimental data and results from other force fields. An increase in the degree of ion dissociation with higher $x_{\mathrm{EC}}$ is observed, consistent with experimental trends.\cite{Binary_Exp} APPLE\&P predicted both $\alpha_{\mathrm{s}}$ and $\alpha_{\mathrm{d}}$ approaching zero at low $x_{\mathrm{EC}}$ but produced relatively lower values of $\alpha_{\mathrm{s}}$ (or $\alpha_{\mathrm{d}}$) at higher $x_{\mathrm{EC}}$.\cite{APPLEP_2009} In contrast, the Class-II force field predicted largely dissociated Li ions even at $x_{\mathrm{EC}} =$ 0.1,\cite{Ravikumar} which significantly disagrees with experimental data. (To note, the results in refs~\citenum{APPLEP_2009} and \citenum{Ravikumar} were obtained at a salt concentration of 1 M, not 1 mol/kg. Here, 1 mol/kg of \ce{LiPF6} corresponds to 1.02–1.16 M, depending on $x_{\mathrm{EC}}$.)
The observed dissociation trend is consistent with the decreasing coordination of PF$_6^-$ ions with increasing $x_{\mathrm{EC}}$ in Fig.~\ref{fgr:Solvation}a. 

In Fig.~S8b\dag\ and the accompanying text, we conducted a similar analysis of the degree of ion dissociation for EC, PC, DMC, and DEC solvents with 1 M \ce{LiPF6} salts. The dominance of \ce{LiPF6} ion pairs in solvents with low dielectric constants, such as DMC and DEC, and their dissociation in high dielectric constant solvents, such as EC and PC, are consistent with the results for EC/DMC mixtures and align with infrared spectroscopy analysis.\cite{LiPF6_IR} These results demonstrate that the pretrained potential effectively captures the dynamic variations in dielectric shielding between cyclic and linear solvents.

\subsubsection{Diffusivities}
The diffusivity of anions and cations in electrolytes is an important property that determines battery performance. We theoretically obtained the self-diffusion coefficient ($D$) for \ce{Li+} and 
PF$_6^-$ in pure solvents like PC and DMC as well as mixed solvents of EC/DMC, for which the experimental data are available. As in the previous subsection, we employed the experimental densities, while results with the theoretical densities are also discussed. 

For single solvent electrolyte systems, simulations were performed in the NVT ensemble, following a procedure largely similar to that used for the EC/DMC 1 mol/kg \ce{LiPF6} system described in the previous subsection. The temperatures were set according to where the experimental diffusivities were measured: 293 K for PC or DEC\cite{ECPCDEC_1M_LiPF6} and 298 K for DMC.\cite{DMC_1M_LiPF6} Experimental densities were used to generate initial configurations,\cite{Exp_D2, SigmaAldrich_DMC1M, SigmaAldrich_DEC1M} and the numbers of solvent and salt molecules were adjusted to achieve a 1 M \ce{LiPF6} concentration with total number of atoms close to 1000. Detailed information on the number of molecules and the simulation box lengths are listed in Table S7.\dag\ An equilibration run of 1.4 ns was followed by a 1-ns production run, similar to the EC/DMC 1 mol/kg \ce{LiPF6} simulation. Snapshots were sampled every 100 fs for calculation of ion diffusivity. Five independent runs were conducted with different initial configurations for statistical average.

For the EC/DMC 1 mol/kg \ce{LiPF6} binary solvent electrolyte system, longer production runs were 
found to be essential for diffusion analysis, likely due to the more complex solvation nature of ions in binary solvents compared to single solvent electrolyte systems. Starting from the NVT MD simulations described in the former subsection, four compositions were selected, and the production runs for each composition were extended to 7 ns across three independent runs to obtain reliable diffusivity data.

From the MD trajectories during the production run, the self-diffusion coefficient was calculated by the mean squared displacement (MSD) for each ion type. In detail, with a given time window $\tau$ and position vector $\mathbf{r}_i(t)$ of a particle $i$ at time $t$, the squared displacement was averaged over all particles with the same type and all available time origins:\cite{10.1038/s41524-018-0074-y}
\begin{align}
\label{eqn:MSD}
\mathrm{MSD}(\tau)
& = \langle  |\mathbf{r}_i(t+\tau)-\mathbf{r}_i(t)|^2 \rangle_{t, i} \\
& = \frac{1}{N_{\tau}} \sum_{t=0}^{T-\tau} \frac{1}{N_{i}} \sum_{i=0}^{N} |\mathbf{r}_i(t+\tau)-\mathbf{r}_i(t)|^2
\end{align}
where $N_{\tau}$ is the number of available time windows and $N_{i}$ is the number of particles of interest. We then obtained $D$ using the Einstein relation.\cite{Textbook_Allen}
\begin{equation}
D = \lim_{t\to\infty}\frac{\mathrm{MSD}(t)}{6t}
\end{equation}

The linear regression of an MSD--$\tau$ curve using appropriate bounds efficiently captures the linear region of diffusion.\cite{10.1038/s41524-018-0074-y} We set the lower and upper bound as 10\% and 60\% of the total production time, respectively, ensuring accurate identification of the linear region in the MSD--$\tau$ curve. 

\begin{figure*}
\centering
  \includegraphics[width=17.1cm]{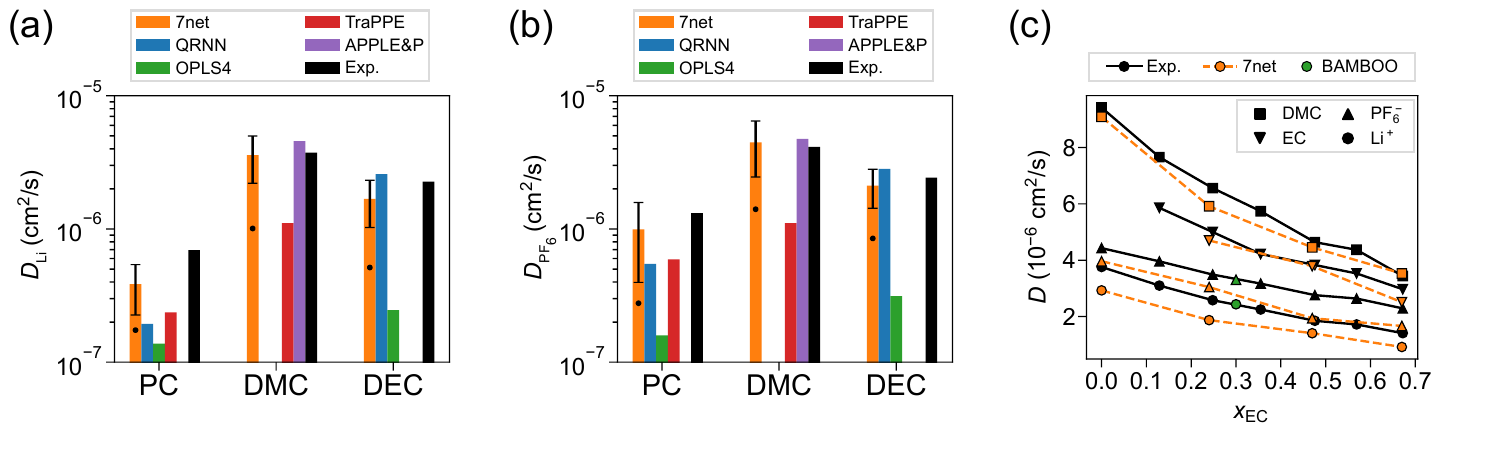}
  \caption{Comparison of diffusivity of (a) Li and (b) PF$_6$  obtained using SevenNet at experimental density (orange), QRNN\cite{QRNN} (blue), OPLS4\cite{QRNN} (green), TraPPE\cite{TraPPE} (red), APPLE\&P\cite{APPLEP_2009} (purple), and experiments\cite{Exp_D1, Exp_D2} (black). Error bars indicate the standard deviation across five simulations. The black dots on the orange bar  indicate the calculated diffusivity at the equilibrium density obtained from SevenNet-NPT simulations. (c) Diffusivity of solvents and ions in EC/DMC (1 mol/kg \ce{LiPF6}) electrolyte at 298 K. Values were averaged over 3 simulations. Experimental values were adopted from ref~\citenum{Binary_Exp}. The results by BAMBOO are also displayed.\cite{BAMBOO}
  }
  \label{fgr:diffusivity}
\end{figure*}

In Fig.~\ref{fgr:diffusivity}a and b, we present the calculated diffusivities of Li$^+$ and PF$_6^-$ ($D_{\rm Li}$ and $D_{\rm PF_6}$, respectively) when 1 M \ce{LiPF6} is desolved in PC, DMC, or DEC solvents. The corresponding MSD--$\tau$ curves with the $R^2$ values obtained from the linear regression are presented in Fig.~S11.\dag\ For comparison, other computational\cite{QRNN, TraPPE, APPLEP_2009} and experimental\cite{Exp_D1, Exp_D2} results are also provided.  The actual diffusivity values, along with predicted and experimental densities, can be found in Table S8.\dag\  
Among the computational approaches, SevenNet achieved reasonable accuracy in a consistent way. 
In comparison, other computational methods showed larger discrepancies with experiment although they accurately predicted pure solvent density. This might be attributed to the general overestimation of viscosity in the QRNN and OPLS4 models and the overestimation of relative permittivity in the TraPPE model. The former slows molecular movement, while the latter affects solvent dynamics. In contrast, the APPLE\&P model demonstrated less than 4\% error in the density of 1 M \ce{LiPF6} in the DMC solvent and achieved accuracy comparable to SevenNet. SevenNet also successfully captured the general trend of higher anion diffusivities compared to cations, consistent with previous studies.\cite{AIMD_EC_EMC} This phenomenon likely results from the strong solvation of Li ions by the solvents, which hinders their mobility relative to PF$_6^-$ ions. Consistently, the higher permittivity of PC enhances solvation and suppresses Li-ion diffusivity relative to the DMC solvent.
In Fig.~\ref{fgr:diffusivity}a and b, the diffusivities obtained when the SevenNet equilibrium density was used are marked in black dots in the middle of the bar. The diffusivity was significantly reduced by 50\%$\sim$70\%, underscoring the critical role of density. 

In Fig.~\ref{fgr:diffusivity}c, the diffusivity of ions in the EC/DMC (1 mol/kg \ce{LiPF6}) binary solvent system shows good agreement with experimental values, albeit with slight underestimation.\cite{Binary_Exp} The overall decrease in diffusivity with increasing $x_{\mathrm{EC}}$ reflects reduced electrolyte mobility as the concentration of the highly viscous EC solvent increases, indicating that the variation in viscosity is also well described by SevenNet. Notably, the general trend of $D_{\textrm{DMC}} > D_{\textrm{EC}} > D_{\ce{PF6}} > D_{\ce{Li}}$ is consistent across all compositions. The lowest diffusivity of \ce{Li+} suggests its largest hydrodynamic size, even greater than that of the \ce{PF6-} anion, due to strong coordination with surrounding solvent molecules.\cite{Binary_Exp}

\subsection{Analysis on the training dataset}
In the preceding subsections, the overall performance of SevenNet in simulating liquid electrolytes was found to be satisfactory. To investigate whether this accuracy arises from adequate sampling in the training set, we analyzed the Materials Project Trajectory (MPtrj) dataset\cite{CHGNet} for the presence of relevant solvent molecules and chemical moieties.
We first identified molecular units embedded in the inorganic compounds that include O, C, and H atoms. The detailed procedure is described in the text accompanying Fig.~S12.\dag\ Among the 20 solvent molecules studied in this work, only the DME molecule was found in four compounds (Fig.~S12a\dag). To explore the presence of similar types of molecules, we manually inspected molecular units consisting of 6–24 atoms, including O, C, and H, and identified similar motifs, such as ether, ester, and five-membered ring groups (Fig.~S12b\dag). However, no carbonate groups were found.

We further examined the presence of the local chemical moieties classified in Fig.~S1.\dag\  As shown in Fig.~S13,\dag\ structures containing species such as \ce{O_c}/\ce{C_c}, \ce{C=C}, and \ce{C_{CH3}} were prevalent in the training set. This abundance of these chemical moieties can be understood by general chemistry.
However, several moieties were scarce in the training set, in particular fluorinated ones. 
For example, only 32 structures contained the \ce{C_{OFH2}} moiety, and the \ce{C_{CFH2}} moiety was absent. 
Such data scarcity may explain the pronounced softening for the fluorine atoms in Fig.~\ref{fgr:Intramolecular}c. 

Extending the analysis to the Li solvation shell, we identified structures in the MPtrj training set containing \ce{Li-O_c} or \ce{Li-O_e}, as well as oxygen atoms bonded to \ce{Na+} and \ce{K+}, which share similar chemical properties with \ce{Li+}.
Specifically, we found 183 and 63 structures containing \ce{Li-O_c} and \ce{Li-O_e} moieties, respectively (see Fig.~S14\dag\ for representative structures). The \ce{Li-O_c} structures predominantly included carbonate ions and their derivatives. Notably, a substantial number of these structures contained transition metals coexisting with \ce{Li+}, \ce{Na+}, and \ce{K+}, which are commonly used in LIB cathodes. The exhaustive search for new cathode materials has likely contributed these structures to the databases. The Li ions are mostly charged in the corresponding structures, which allowed SevenNet to learn Coulomb interactions between \ce{Li+} and \ce{O_c} or \ce{O_e}.  

The foregoing analysis indicates that significant portions of molecular units or chemical moieties were directly sampled in the MPtrj dataset, suggesting that relevant local chemical bondings were mostly captured. However, non-local bondings that contribute to the formation of whole molecular units and intermolecular interactions, in particular between organic molecules, were not sufficiently sampled.
This suggests that the model learned large parts of the PES by generalizing across the chemical space, facilitated by deep learning and learnable atomic embeddings.

To understand model generalization in terms of latent space, we paid attention to \ce{C_{OF2H}} and Li solvation shells. The \ce{C_{OF2H}} moiety was absent in the training set but exhibited a notable degree of force accuracy (see Fig.~\ref{fgr:Intramolecular}c).
This means that the model interpolate or extrapolate from other structures in the training set. For Li solvation shells, while the training set included structures containing \ce{Li-O_c} and \ce{Li-O_e} moieties, it remained uncertain how the model handled multiple \ce{Li-O} interactions within solvation shells.

For both test cases, we analyzed the atomic descriptors in the latent space to examine relative proximity or similarity between structures. We first extracted 128-dimensional invariant atomic descriptors, which served as input vectors for the output block producing atomic energies. Dimensionality reduction techniques, including Principal Component Analysis (PCA) and Uniform Manifold Approximation and Projection (UMAP),\cite{UMAP} were employed. 
Using PCA, we reduced the dimensionality to 31 components, retaining 95\% of the original variance, and applied a whitening scheme. As shown in Fig.~S15a,\dag\ SevenNet positioned the \ce{C_{OF2H}} moiety—absent in the training set—between the \ce{C_{OFH2}} and \ce{C_{OF3}} moieties. Furthermore, as the \ce{F}:\ce{H} ratio shifted from 3:0 to 0:3, the PCA data points for each moiety aligned linearly. That is to say, SevenNet interpolated untrained regions by leveraging knowledge derived from trained regions.

Next, we performed UMAP analysis\cite{SevenNet-UMAP} on both the training set and sampled structures from the simulation on the Li solvation shells. Euclidean distances between atomic descriptors were used to construct a high-dimensional graph, where similar descriptors were positioned closer together, and dissimilar descriptors were placed farther apart.
In particular, we examined distinct solvation types identified in Section 2.3.1 by extracting MD trajectories (Fig.~S15b\dag) for separate analysis. For each solvation type, the structure with the minimum potential energy among all snapshots was selected. The analysis in Fig.~S15b\dag\ revealed that the Li environments within solvation shells formed distinct clusters in the latent space. The structures in the training set interpreted as being similar to these Li environments typically featured multiple oxygen atoms bonded to either Li or Na. While the training set did not include the same Li--solvent structures encountered during MD simulations, the model effectively learned first-neighbor interactions from these examples, where multiple oxygen atoms are bonded to a Li ion. 

The above analysis on the latent space indicates that SevenNet might infer untrained regions by generalizing knowledge from trained regions. However, full understanding of the generalization is beyond the current scope because of the black-box nature of deep learning models. 

\subsection{Fine-tuning the pretrained model}
In the previous sections, we have demonstrated the capabilities and limitations of SevenNet on liquid electrolytes. Intermolecular interactions, particularly liquid density, are critical to the physicochemical properties of liquid electrolytes. However, SevenNet was less accurate for the intermolecular interactions, leading to the overestimation in the liquid density and underestimation of diffusivities. It has been shown that fine-tuning pretrained models can achieve accuracies comparable to bespoke models.\cite{PerformanceAssessment, Sublimation, SystematicSoftening}
As an example of fine-tuning in the present applications, we selected the DMC solvent, which exhibited significant overestimations of density in Fig.~\ref{fgr:pure_density}a. 

To construct the training set for fine-tuning, we conducted MD simulations using SevenNet for 100 ps in the 298 K NVT ensemble with 360 atoms and the experimental density. Subsequently, DFT single-point calculations were performed on 50 snapshots extracted from the last 50 ps at intervals of 1 ps. These 50 structures were further modified by scaling the lattice parameters, while maintaining fixed intramolecular distances, by factors of 0.9 and 1.1.\cite{Csanyi} This procedure generated a total of 150 structures for the fine-tuning training set. 

We fine-tuned the model using the same parameters as SevenNet, with adjustments to the learning rate and stress loss weight. The learning rate started at $10^{-4}$ and decreased linearly to $10^{-6}$ over 600 epochs. Furthermore, the weight of the stress loss was increased from 0.01 to 1.0 to enhance the fine-tuning effect on solvent density. To monitor knowledge retention from the original SevenNet model, we evaluated the MAEs on a test set of 19 072 structures containing \ce{O}, \ce{C}, and \ce{H}, filtered from the SevenNet training set (see Fig.~S16\dag). The fine-tuned model (hereafter referred to as SevenNet-FT) after 50 epochs, achieved MAEs of 0.032 eV/atom, 0.086 eV/Å, and 0.57 kbar for energy, forces, and stresses, respectively. Notably, the stress MAE decreased from 2.78 kbar (SevenNet) to 0.57 kbar (SevenNet-FT). In terms of computational cost, the entire fine-tuning procedure, including training set generation, required only a few hours on a moderate computing node.

Using SevenNet-FT, we obtained the liquid densities following the procedure described in Section 2.3.1, and the results are shown in Fig.~\ref{fgr:pure_density}a as x-markers. The liquid densities of linear carbonates align well with experimental results, whereas those of other solvents are underestimated. The normal stress parity plots for pure solvents (Fig.~S17\dag) reveal significant improvements for linear carbonates, moderate improvements for cyclic carbonates, ethers, and esters, but only minor improvements for solvents containing fluorine atoms. This behavior may be attributed to the fine-tuning training set, which includes only DMC molecules composed of O, C, and H atoms, thereby limiting its applicability to cyclic carbonates and fluorinated systems. Additionally, forgetting of knowledge from SevenNet was not fully prevented during the fine-tuning process (Fig.~S16\dag), which may have contributed to the underestimation of densities for PC, FEC, and DFEC.

\section{Summary and Conclusion}
In summary, we applied a pretrained universal interatomic potential, SevenNet, to the simulations of liquid electrolytes in LIBs. 
Even though SevenNet was mostly trained on the inorganic compounds, it demonstrated sound predictive capabilities for key properties such as solvation structures and diffusivities. However, it also exhibited limitations, particularly in predicting liquid density. These limitations primarily stem from the lack of explicit representation of organic systems and intermolecular interactions in the original training set, which was largely focused on inorganic compounds.
Despite these challenges, the model’s ability to generalize across chemical spaces improved the accuracy in the strongly out-of-distribution domains. 
Analysis of latent space suggested that SevenNet leverages learnings from related chemical moieties and structural motifs to interpolate and predict untrained regions.
Fine-tuning SevenNet for specific cases, as demonstrated with DMC solvents, significantly improved accuracy in density and stress predictions, with minimal computational costs. This approach paves the way for tailoring pretrained models to specialized applications, making them useful tools for material discovery and optimization in electrolyte engineering.
Another potential approach would be to train a model using both an inorganic crystal database and a molecular database\cite{SPICE} through multi-fidelity training,\cite{Multifidelity} which will be explored in a future study.
In conclusion, this work underscores the potential of SevenNet for advancing the engineering of liquid electrolyte systems, thereby accelerating the development of next-generation LIBs.

\section{Methods}

\subsection{DFT calculation}
All DFT calculations in this work were performed using the Vienna \textit{ab initio} simulation package (\texttt{VASP}), employing the projector-augmented wave (PAW) pseudopotentials.\cite{10.1103/physrevb.59.1758} The Perdew-Burke-Ernzerhof (PBE) exchange-correlation functional, based on the generalized gradient approximation (GGA), was used for electrons.\cite{10.1103/physrevlett.77.3865}
For condensed phases of organic molecules, the van der Waals dispersion interaction plays an important role in determining quantities such as density and viscosity. Since semilocal functionals such as PBE do not take into account the dispersion interactions, we added the Grimme's D3 dispersion correction with Becke-Johnson (BJ) damping.\cite{10.1063/1.3382344, 10.1002/jcc.21759} 
The dispersion and coordination cutoff radii in the D3 correction term were set to 50.2 and 20.0 Å, respectively.
The PBE-D3 functional shows similar accuracy to the PBE0-D3 (hybrid GGA) functional in predicting dimer interaction energies, while also achieving density predictions for organic crystals\cite{D3_organic} and ionic liquids\cite{D3_ionic} that fall within 1\% of experimental values. For molecules and dimers in vacuum, spin-polarized calculations were performed with a cutoff energy of 520 eV, while for bulk liquid configuration, a spin-unpolarized setting was used.
However, we found that the spin-polarization was negligible in all isolated molecules.
The Brillouin zone was sampled only at the $\Gamma$-point. Concerning the specific PAW pseudopotentials, those without suffixes were used except for Li\_sv, in alignment with the calculation settings in Materials Project.

\subsection{Pretrained model}
In this work, we utilized a pretrained model SevenNet-0 (version 11July2024),\cite{SevenNet, SevenNet-0} which is based on the architecture of NequIP.\cite{NequIP} SevenNet has achieved high performance in the Matbench Discovery benchmark, which assesses the performance of pretrained universal force fields on inorganic crystal discovery.\cite{Matbench} As a GNN-IP, SevenNet initializes node and edge features from atomic numbers and relative position vectors, respectively. An edge connects two nodes if their interatomic distance is less than a pre-defined cutoff radius.
Starting from these features, multiple message-passing layers aggregate information from connected nodes and edges. After the last message-passing layer updates node features, they are used to predict total energy by the readout layer. Forces and stresses are derived from the energy gradient.
Although the SevenNet model hyperparameters remain identical to its previous version in ref~\citenum{SevenNet}, the training dataset has been updated to the MPtrj dataset without dataset splitting. The learning rate was initialized at 0.01 and decreased linearly to 0.0001 over 600 epochs. As a result, SevenNet achieves the MAEs of 0.011 eV/atom, 0.041 eV/Å, and 2.78 kbar for energy, forces, and stresses, respectively.

We employed the Atomic Simulation Environment (\texttt{ASE}) interface\cite{ASE} and the Large-scale Atomic/Molecular Massively Parallel Simulator (\texttt{LAMMPS}) package,\cite{LAMMPS} in conjunction with SevenNet, to compute the structural and dynamical properties of solvents and electrolytes.
The \texttt{ASE} interface was used for static calculations such as (constrained) relaxation while the \texttt{LAMMPS} package was used for MD simulations.
To account for dispersion interactions absent in the MPtrj dataset, we integrated an in-house CUDA implementation of Grimme's D3 dispersion correction with BJ damping\cite{10.1063/1.3382344, 10.1002/jcc.21759} into SevenNet, ensuring that the calculations were performed at the PBE-D3 level of theory, which is identical to the DFT calculation in the previous section. The parameter set for the dispersion interaction was consistent with that used in the DFT calculations (see above).

\section*{Author contributions}
S. Ju and J. You conducted the DFT calculations, simulations using the pretrained model, data analyses, and script writing. G. Kim and H. An developed the CUDA-accelerated D3 implementation for LAMMPS. Y. Park generated SevenNet-0 and provided guidance on potential-related scripting. S. Han supervised all aspects of the work, including DFT calculations, simulations, and script development. All authors reviewed, provided feedback on, and approved the final manuscript.

\section*{Conflicts of interest}
There are no conflicts to declare.

\section*{Data availability}
The data supporting this article have been included as part of the ESI$\dag$ of this article.

\section*{Acknowledgements}

This work was supported by the National Research Foundation of Korea (NRF) grant funded by the Korea government (MSIT) (RS-2023-00247245). The computations were carried out at the Korea Institute of Science and Technology Information (KISTI) National Supercomputing Center (KSC-2024-CRE-0215) and at the Center for Advanced Computations (CAC) at Korea Institute for Advanced Study (KIAS).


\balance

\bibliography{reference}
\bibliographystyle{reference}
\end{document}